\documentclass[12pt]{article}

\usepackage{epsfig}
\usepackage{amssymb}
\usepackage{graphics}
\let\a=\alpha   \let\b=\beta   \let\g=\gamma

           \let\p=\pi      
  \let\t=\tau

\def\ie{{\it i.e.\ }}
\newcommand{\be}{\begin{equation}}
\newcommand{\ee}{\end{equation}}
\newcommand{\bea}{\begin{eqnarray}}
\newcommand{\eea}{\end{eqnarray}}
\newcommand{\ba}{\begin{array}}
\newcommand{\ea}{\end{array}}
\def\nn{\nonumber}

\begin{document}
%
%
%
%
\begin{titlepage}
\rightline{hep-th/0503061}
%
%
\vskip 2cm
\centerline{{\large\bf Classification and a toolbox for
orientifold models}}
\vskip 0.3cm
%
%
\vskip 1cm
\centerline{
P. Anastasopoulos\footnote{panasta@physics.uoc.gr}$^{,\a,\b}$,
A. B. Hammou\footnote{amine@physics.uoc.gr}$^{,\a,\g}$}
\vskip 1cm
\centerline{$^\a$ Department of Physics, University of Crete,}
\centerline{71003 Heraklion, GREECE.}
\vskip  .5cm
\smallskip
\centerline{$^\b$ Laboratoire de Physique Th\'eorique Ecole
Polytechnique,} \centerline{91128 Palaiseau, FRANCE.}
\vskip  .5cm
\smallskip
\centerline{$^\g$
%
%
%
Departement de Physique, Universit\'e des Sciences et Technologie
d'Oran,} \centerline{BP 1505, Oran, El M'Naouer, ALGERIE}
\vskip 1cm
\begin{abstract}
We provide the general tadpole conditions for a class of
supersymmetric orientifold models by studing the general
properties of the elements included in the orientifold group. In
this talk, we concentrate on orientifold models of the type
$T^6/Z_M\times Z_N$.
\end{abstract}
\end{titlepage}

\section{Introduction}

Orientifolds are a generalization of orbifolds \cite{sagn,
Pradisi:1988xd, horava}, where the orbifold symmetry includes
orientation reversal on the worldsheet. The orientifold group
contains elements of two kinds: internal symmetries of the
worldsheet theory forming a group $G_1$ and elements of the form
$\Omega\cdot g$, where $\Omega$ is the worldsheet parity
transformation and $g$ is some symmetry element now taken from a
group $G_2$. Closure implies that $\Omega\cdot g\cdot \Omega\cdot
g^{\prime} \in G_1$ for $g, g^{\prime} \in G_2$. The full
orientifold group is $G_1+\Omega G_2$. The one loop amplitude
implementing the $\Omega$ projection is interpreted as Klein
Bottle amplitude and has in general ultraviolet divergences
(tadpoles). These ultraviolet divergences are interpreted as
sources in space time, that couple to the massless type IIB
fields, the metric, the dilaton and the RR-forms. They are
localized in sub-manifolds of space-time, known as Orientifold
planes, $O_p$. These are non dynamical objects characterized by
their charges and tensions. Consistency and stability of the
theory are assured if D-branes are introduced in a way that
guarantees the cancellation of these tadpoles. RR-tadpole
cancellation is equivalent to the vanishing of gauge charge in a
compact space, whereas NS-tadpole cancellation is equivalent to
the vanishing of forces in the D-brane/O-plane vacuum
configuration. In this talk we study the tadpole conditions for a
class of orientifold groups of the type $G+\Omega ~G$ where $G$ an
orbifold group. Consider the projection of type IIB string theory
on $R^4\times T^6$ by the orientifold group $G+\Omega~G$. The
orbifold group $G$ acts on the coordinates $z^i$ of the
6-dimensional torus $T^6$ that we take to be factorized as
$T_1^2\times T_2^2 \times T_3^2$ as
%
\be \a = e^{2\pi i \sum_{i=1}^3 (v_\a^i J^i + r_i~\delta_\a^i
~P^i)} , \nonumber \ee
with $J^i$ and $P^i$ the generators of rotations and diagonal
translation in each of the internal two torus $T_i^2$ with radius
$r_i$. To preserve supersymmetry the $v_\a^i$'s should satisfy the
condition $\sum_i v_\a^i=0$.

In this talk we restrict ourself to the case of rotations
preserving some supersymmetry, and refer to \cite{ourpaper} for
the non supersymmetric models including shifts. It is not
difficult to be convinced that the most general rotation elements
$\a$ preserving some supersymmetry are such that
$v_\a=(v_\a^1,v_\a^2,v_\a^3)$ with say $v_\a^3=0$ or $v_\a^3\neq
0$.

\section{Supersymmetric Orientifolds}

\begin{figure}
\begin{center}
\epsfig{file=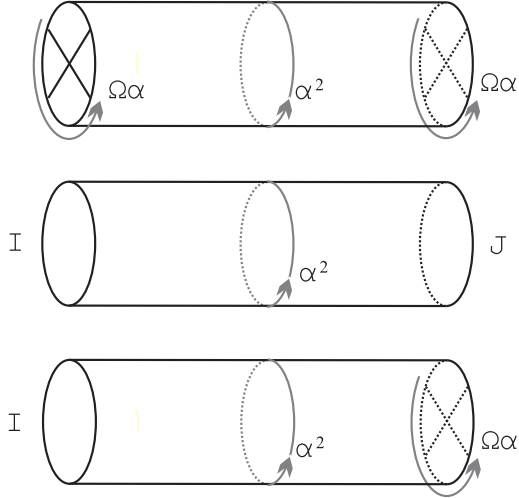,width=70mm}
\end{center}
\caption{Klein-bottle, Annulus and M\"obius strip. The one-loop
amplitudes become tree-level in the transverse picture where an
$\a^2$-twisted closed string propagates between crosscaps and
boundaries.}\label{KleinAnnulusMobius}
\end{figure}

\subsection*{Klein Bottle}

Consider an element $\a$ of $G$. We can work out the contribution
of this element to the Klein Bottle amplitude by using the trace
formula:
\bea {\cal K}_\a = Tr_{U+T}\left[ \Omega \a ~~q^{L_0}
\bar{q}^{\bar{L}_0}\right], \nonumber \eea
where the trace is over untwisted $U$ and twisted $T$ closed
string states of the type IIB orbifold model considered. Due to
the presence of $\Omega$ the only states contributing to ${\cal
K}_\a$ are the untwisted states and the $Z_2$ twisted ones (by
$Z_2$ we really mean order two elements i.e. $R^2=1$). The later
exist only if the orbifold group $G$ contains $Z_2$ factors.
%
%
%
%
%
%
%
%
%
%
%
%
%

The contribution of an element $\a$ corresponds to a propagation
of a closed string state projected out by $(\Omega \a)^2=\a^2$ and
$(\Omega R \a)^2= \a^2$ when $R\in G$ since all elements of $G$
commute with $\Omega$ in our case.

%
%
To extract the massless tadpole contribution we perform a modular
transformation $l=1/4t$ where $t$ is the loop modulus and $l$ the
cylinder length \cite{Gimon:1996rq, Gimon:1996ay} and then take
the limit $l\to \infty$.
%
%

%
%

The general contribution to the tadpoles from Klein-Bottle
amplitudes of a group element $\a$ such that
$v_\a=(v_\a^1,v_\a^2,0)$ is:
\bea && (1_{NS}-1_{R})\frac{1}{\prod_l 2\sin2\pi v_\a^l}
\Bigg\{\sqrt{{\cal V}_3} \bigg(\prod_{l} 2\cos\pi v_\a^l +
\prod_{l} 2\sin\pi v_\a^l\bigg)
\nonumber\\
&&\qquad \qquad ~~~ - \frac{1}{\sqrt{{\cal V}_3}}\sum_{i\neq
j=1,2} \epsilon_{ij}~2\cos\pi v_\a^i 2\sin\pi v_\a^j \Bigg\}^2~.
\label{kiii} \eea
where $\epsilon_{12}=-\epsilon_{21}=1$. The first term corresponds
to the contribution of $\Omega \a$, the second term is due to
$\Omega R_3 \a$ and both are proportional to the volume ${\cal
V}_3$ of the third torus\footnote{By $R_i$ we denote a $Z_2$
rotation element that acts by reflecting the coordinates of the
tori $T^j$ with $i \neq j$.}. The remaining terms are the
contribution from $\Omega R_i \a$ and are proportional to the
inverse volume. The amplitude above is proportional to
$(1_{NS}-1_{R})$ and appears as perfect square
\cite{Pradisi:1988xd,Angelantonj:2002ct}.
%

Consider now $v_\a=(v_\a^1,v_\a^2,v_\a^3)$ where
$v_\a^{l=1,2,3}\neq 0$.
\bea (1_{NS}-1_{R}) \frac{1}{\prod_l 2\sin2\pi v_\a^l}
\bigg(\prod_l 2\cos\pi v_\a^l + \sum_i 2\cos\pi v_\a^i
\prod_{l\neq i} 2\sin\pi v_\a^l\bigg)^2 \label{kkii} \eea
The amplitude is again perfect square as it should
\cite{Pradisi:1988xd,Angelantonj:2002ct}. The different terms
correspond as before to $\Omega \a$ and $\Omega R_l \a$ with
$l=1,2,3$ respectively for $R_l\in G$.
%

In orientifold models tadpoles arise as divergences at the
one-loop level, which are interpreted as inconsistencies in the
field equations for the R-R potentials in the theory. These
tadpoles can be regarded as the emission of an R-R closed string
state from a Dp-brane (disc), a source of (p+1)-form R-R potential
and from an orientifold plane (${\bf RP}^2$), which carries R-R
charges. The contributions we obtained above come from $Op$-planes
sources of twisted (p+1)-form R-R potentials. For non trivial
twists, the string zero modes vanish, and therefore the $Dp$-brane
is forced to sit at the fixed point. It simply means that the
twisted R-R potential can not propagate in peacetime.

\subsection*{Annulus}

To cancel the aforementioned Klein-Bottle UV divergences
-tadpoles- we need to add D-branes to the spectrum (open string
sector). For the model under consideration these are a bunch of
D9-branes and D5$_i$-branes extended along the $T_i^2$ torus and
siting on the $R_i$-fixed points when the group $G$ contains
$R_i$-factors. The Annulus amplitudes can be easily computed
between all kinds of D-branes existing in the theory, with the
contribution of the group element $\a$ given by the trace formula:
\bea {\cal A}_{IJ,\a} = Tr_{IJ} \left[\a ~q^{L_0} \right]~,
\nonumber \eea
where now the trace is over all open string states attached
between $I$ and $J$ D-branes for $I,J=9,5_i$ siting at $\a$-fixed
point. For simplicity we will always put the branes at one fixed
point. To extract the tadpole contributions we need to perform a
modular transformation to the transverse channel $l=1/2t$ and then
take the limit $l\to \infty$ \cite{Gimon:1996rq}.
%
%

For $\a$ such that: $v_\a=(v_\a^1,v_\a^2,0)$ we have the following
contribution to the tadpoles:
\bea && (1_{NS}-1_{R}) \frac{1}{\prod_l 2\sin2\pi v_\a^l}
\Bigg[\sqrt{{\cal V}_3} \Big(Tr[\gamma_{\a,9}]+ \prod_{l} 2\sin\pi
v_\a^l Tr[\gamma_{\a,5_3}]\Big)
\nn\\
&&\qquad \qquad ~~~ - \frac{1}{\sqrt{{\cal V}_3}}\sum_{i\neq
j=1,2} 2\sin\pi v_\a^j Tr[\gamma_{\a,5_i}]\Bigg]^2. \label{aiii}
\eea
The different terms correspond are the contribution to the
tadpoles from the D-branes existing in the theory, and $\g_{\a,
I}$ is the action of the group element $\a$ on the Chan-Paton
factors in the D$I$-brane sector. The amplitudes is again
proportional to zero $(1_{NS}-1_{R})$ reflecting the fact that the
orientifold group action preserve some supersymmetry.
%

For $v_\a=(v_\a^1,v_\a^2,v_\a^3)$ the contribution to the tadpoles
from the Annulus amplitudes is as follows:
\bea (1_{NS}-1_{R}) \frac{1}{\prod_{l} 2\sin2\pi v_\a^l}
\Big(Tr[\gamma_{\a,9}]+ \sum_{i=1}^3 \prod_{l\neq i} 2\sin\pi
v_\a^l Tr[\gamma_{\a,5_i}]\Big)^2. \label{aaii} \eea
The structure of these amplitudes is similar to (\ref{aiii})
without the volume dependance and with the product extended over
$l=1,2,3$. It is explicitly perfect square as expected
\cite{Pradisi:1988xd,Angelantonj:2002ct}.
%
%

\subsection*{M\"obius Strip}

Finally, the contribution of the group element $\a$ to the
M\"obius strip amplitude can be computed from the trace formula
\bea {\cal M}_{I,\a} = Tr_I \left[\Omega \a ~q^{L_0}\right], \eea
where the trace is over open strings attached to D$I$-branes. To
extract the contribution to the tadpoles we must perform a modular
transformation to the transverse channel by $P=T S T^2 S T$ where
$T:\t \to \t+1$ and $S:\t \to - 1/\t$, $l=1/8t$. Finally, we take
the UV limit $l \to\infty$. The M\"obius strip transverse channel
amplitude is the mean value of the transverse channel Klein Bottle
and Annulus amplitudes \cite{Pradisi:1988xd, Angelantonj:2002ct}.
Taking into account this fact, extracting the UV limit in the
M\"obius strip amplitude and comparing with the amplitudes
obtained in the Klein Bottle and Annulus gives constraints on the
matrices $\g_{\a,I}$ and $\g_{\Omega.\a,I}$:
\bea &&Tr[\gamma^T_{\Omega \a,9}\gamma^{-1}_{\Omega
\a,9}]=Tr[\gamma_{\a^2,9}],
\nonumber\\
&&Tr[\gamma^T_{\Omega R_i \a,9}\gamma^{-1}_{\Omega R_i \a,9}]=
-Tr[\gamma_{\a^2,9}],
\nonumber\\
&&Tr[\gamma^T_{\Omega \a,5_i}\gamma^{-1}_{\Omega \a,5_i}]=
-Tr[\gamma_{\a^2,5_i}],
\nonumber\\
&&Tr[\gamma^T_{\Omega R_i \a,5_i}\gamma^{-1}_{\Omega R_i \a,5_i}]=
Tr[\gamma_{\a^2,5_i}],
\nonumber\\
&&Tr[\gamma^T_{\Omega R_j \a,5_i}\gamma^{-1}_{\Omega R_j \a,5_i}]=
-Tr[\gamma_{\a^2,5_i}], \label{consta} \eea
where in the last equation $i\neq j$ and $i,j= 1,2,3$. These
constraints apears to be the same for both
$v_\a=(v_\a^1,v_\a^2,0)$ or $v_\a=(v_\a^1,v_\a^2,v_\a^3)$.

\subsection*{Tadpole conditions:}\label{Tadpoles-NoSS}

The massless part of the transverse channel amplitude $\tilde{\cal
K}_\a +\tilde{\cal A}_\a +\tilde{\cal M}_\a$ gives the tadpole
conditions.
\begin{itemize}
\item $\a$ is such that $v_\a=(v_\a^1,v_\a^2,0)$, then:
\bea Tr[\gamma_{\a^2,9}]+4 \prod_{l} \sin2\pi
v_\a^lTr[\gamma_{\a^2,5_3}] &=& 32~\bigg(\prod_{l} \cos\pi v_\a^l
+ \prod_{l} \sin\pi v_\a^l\bigg)\, ,
\nonumber\\
\sum_{i\neq j =1,2} 2\sin2\pi v_\a^j Tr[\gamma_{\a^2,5_i}]&=&
32~\sum_{i\neq j =1,2} \epsilon_{ij} \cos\pi v_\a^i \sin\pi
v_\a^j. \label{tvk2iii} \eea
The different terms exist when the corresponding $R$ factors do,
as an example, if say $R_3 \notin G$ then the second term on both
sides of the first equation will be absent.
\item $\a$ is such that $v_\a=(v_\a^1,v_\a^2, v_\a^3)$ then: \bea
Tr[\gamma_{\a^2,9}]+ 4\sum_{i=1}^3 \prod_{l\neq i} \sin2\pi v_\a^l
Tr[\gamma_{\a^2,5_i}] = 32~ \bigg(\prod_l \cos\pi v_\a^l + \sum_i
\cos\pi v_\a^i \prod_{l\neq i} \sin\pi v_\a^l\bigg).
\label{tvl2ii} \eea
\end{itemize}
Note that in all cases the tadpole conditions holds for both NS
and R sectors due to supersymmetry. On the other hand the tadpole
condition for group elements which are not the square of some
other element of $G$ (there is no element $\b \in G$ such that $\a
= \b^2$), will receive contribution only from the Annulus
amplitude. When this element is such that $v=(v_\a^1,v_\a^2,0)$ or
$v=g_3+v_\a$, the tadpole conditions will be the same as before,
with zero on the right hand side. For the elements such that
$v=g_i+v_\a$ the tadpole condition is not difficult to work out,
leading to:
\bea Tr[\gamma_{R_i\a,9}]&+& 4\sin\p v_\a^i \cos\p v_\a^j
Tr[\gamma_{R_i\a,5_3}]
\nonumber\\
&+&2\cos\pi v_\a^j Tr[\gamma_{R_i\a,5_i}]+2\sin\p v_\a^i
Tr[\gamma_{R_i\a,5_j}]=0, \label{tgiv} \eea
where $i\neq j =1,2$ and the different terms exist only if the
corresponding $R_i$ factor does.
If $v_\a=(v_\a^1,v_\a^2,v_\a^3)$, the tadpole conditions are the
same as (\ref{tvl2ii}) without the right hand side (\ie the right
hand side is zero).

\subsection*{Some specific examples}

Let us discuss some of the examples that can be described by the
general formulas we have obtained in the previous sections. The
first example is the groups studied by Gimon and Johnson
\cite{Gimon:1996ay}, where $G=Z_N$ for $N=2,3,4,6$ acting on
$T^4$. The tadpole conditions are given by
(\ref{tvk2iii}-\ref{tvl2ii}) with
$v_\a^1=-v_\a^2=\frac{k}{N},~v_\a^3=0$ leading for odd $N$
\bea Tr[\gamma_{2k,9}]= 32 \cos^2\frac{k}{N}\pi \nonumber \eea
whereas for even $N$:
\bea
&&Tr[\gamma_{2k,9}]-4~\sin^2\frac{2k}{N}\pi~Tr[\gamma_{2k,5_{3}}]=
32~ \cos\frac{2k}{N}\pi
\nonumber\\
&&Tr[\gamma_{2k-1,9}]-4~\sin^2\frac{2k-1}{N}\pi~Tr[\gamma_{2k-1,5_{3}}]=
0~. \nonumber \eea
The tadpole conditions for the groups studied by Zwart
\cite{Zwart:1997aj}, where $G=Z_N\times Z_M$, can be easily
reproduced by using (\ref{tvk2iii}-\ref{tgiv}). As an example the
tadpole conditions for $Z_N\times Z_M$ where both $N$ and $M$ are
odd are given by:
\bea
Tr[\gamma_{2k,0,9}]&=&32~\cos^2\frac{k}{N}\p \nonumber\\
Tr[\gamma_{0,2l,9}]&=&32~\cos^2\frac{l}{M}\p \nonumber\\
Tr[\gamma_{2k,2l,9}]&=& 32~\cos\frac{k}{N}\p~\cos\frac{l}{M}\p~
\cos\bigg(\frac{l}{M}-\frac{k}{N}\bigg)\p. \nonumber \eea
all the other cases can be easily worked out in a similar way.
The tadpole conditions for the groups studied by Ibanez et al
\cite{Aldazabal:1998mr}, where $G=Z_N$ with both $N$ even and odd
and $G=Z_N\times Z_M$ with $N$ or $M$ even can be reproduce using
(\ref{tgiv}) by specifying the vector $v_\a$ for the different
group elements.

\section{Conclusion}

In this talk we have computed the tadpole conditions for the
orientifold projection $G + \Omega G$ of type IIB string theory on
$T^6$ in the supersymmetric case without specifying the orbifold
group $G$. We have found complete agreement with the models
studied in the literature. In this talk we have considered only
the case when the D-branes are sitting on one fixed point mainly
at the origin. However, it is not difficult to study more general
situation where the D5-branes are distributed on different fixed
points. We did not consider the additional freedom of adding
Wilson lines, they are non dynamical. A non-trivial effective
potential is generated for these gauge invariant operators that
dynamically break part of the gauge group. Finally this analysis
can be extended to more general orientifold groups $G_1+\Omega
G_2$ by considering group elements that do not commute with
$\Omega$ as well as asymmetric orbifold groups and to include
fluxes which are T-dual to orientifold models with branes at
angles.

\vskip 1cm

\centerline{\bf\Large Acknowledgments}

The authors would like to thank Elias Kiritsis for very useful
discussions. The work of A. B. Hammou was supported by RTN
contracts HPRN-CT-2000-0131. This work was partially supported by
RTN contracts INTAS grant, 03-51-6346, RTN contracts
MRTN-CT-2004-005104 and MRTN-CT-2004-503369 and by a European
Union Excellence Grant, MEXT-CT-2003-509661.
The work of P. Anastasopoulos was supported by a Herakleitos
graduate fellowship.


\begin{thebibliography}{10}


\bibitem{sagn}
A.~Sagnotti,
arXiv:hep-th/0208020


\bibitem{Pradisi:1988xd}
G.~Pradisi and A.~Sagnotti,
Phys.\ Lett.\ B {\bf 216} (1989) 59.
%


\bibitem{horava}
P.~Horava,
Nucl.\ Phys.\ B {\bf 327} (1989) 461;
Phys.\ Lett.\ B {\bf 231} (1989) 251.
%
D.~Fioravanti, G.~Pradisi and A.~Sagnotti,
Phys.\ Lett.\ B {\bf 321} (1994) 349 [arXiv:hep-th/9311183];


\bibitem{ourpaper}
P. Anastasopoulos, A. B. Hammou, [arXiv:hep-th/0503044].


\bibitem{Gimon:1996rq}
E.~G.~Gimon and J.~Polchinski,
Phys.\ Rev.\ D {\bf 54} (1996) 1667 [arXiv:hep-th/9601038].

\bibitem{Gimon:1996ay}
E.~G.~Gimon and C.~V.~Johnson,
Nucl.\ Phys.\ B {\bf 477} (1996) 715 [arXiv:hep-th/9604129].


\bibitem{Bianchi:1990tb}
M.~Bianchi and A.~Sagnotti,
Nucl.\ Phys.\ B {\bf 361} (1991) 519.

\bibitem{Angelantonj:2002ct}
C.~Angelantonj and A.~Sagnotti,
Phys.\ Rept.\  {\bf 371} (2002) 1 [Erratum-ibid.\  {\bf 376}
(2003) 339] [arXiv:hep-th/0204089].

\bibitem{Zwart:1997aj}
G.~Zwart,
Nucl.\ Phys.\ B {\bf 526} (1998) 378 [arXiv:hep-th/9708040].


\bibitem{Aldazabal:1998mr}
G.~Aldazabal, A.~Font, L.~E.~Ibanez and G.~Violero,
Nucl.\ Phys.\ B {\bf 536} (1998) 29 [arXiv:hep-th/9804026].



\end{thebibliography}
\end{document}